\documentclass[aps,prb,twocolumn,superscriptaddress,oneside,floatfix,amsmath,showpacs,amssymb]{revtex4}

\usepackage{graphicx}
\usepackage{dcolumn}
\usepackage{bm}
\usepackage{calc} 
\usepackage{amsfonts}  
\usepackage{amsmath}

\begin{document}

\title{Nature of ground states in   one-dimensional electron-phonon Hubbard models at half filling}

\author{H. Bakrim} 
\email{hassan.bakrim@USherbrooke.ca }
\author{C. Bourbonnais}
\email{claude.bourbonnais@USherbrooke.ca }
\affiliation{%
Regroupement Qu\'ebecois sur les Mat\'eriaux de Pointe, D\'epartement de physique, Universit\'e de Sherbrooke, Sherbrooke, Qu\'ebec, Canada, J1K-2R1}%

\date{\today}

\begin{abstract}
The renormalization group technique is applied to one-dimensional  electron-phonon Hubbard models  at half-filling and zero temperature. For the Holstein-Hubbard model, the results of one-loop calculations are congruent with the phase diagram obtained by    quantum Monte Carlo simulations in the $(U,g_{\rm ph})$ plane for the phonon-mediated interaction $g_{\rm ph}$ and  the Coulomb interaction $U$. The incursion of  an intermediate phase   between a fully gapped  charge-density-wave state and a Mott antiferromagnet is supported  along with the  growth of its size with  the molecular phonon frequency $\omega_0$.   We find  additional phases enfolding the base boundary of the  intermediate  phase. A  Luttinger liquid  line is found below  some critical $ U^*\approx g^*_{\rm ph}$,  followed at larger $U\sim g_{\rm ph}$ by  a narrow   region  of   bond-order-wave  ordering which is either  charge or spin gapped   depending on  $U$.  For the  Peierls-Hubbard model, the region of the $(U,g_{\rm ph})$ plane with a fully gapped Peierls-bond-order-wave state shows a growing domination over the Mott gapped  antiferromagnet as the  Debye   frequency $\omega_D$ decreases. A power law dependence $g_{\rm ph} \sim U^{2\eta}$ is found to map out the boundary between the two phases,  whose  exponent is in good agreement with the existing quantum Monte Carlo simulations performed  when  a finite nearest-neighbor repulsion term $V$ is added to the Hubbard interaction.  \end{abstract}


\maketitle

\section{Introduction}
 In approaching the physics  of highly correlated electron systems, we often come against   experimental  evidence for   significant coupling between  interacting electrons and   lattice   degrees of freedom of different kinds    \cite{NoteHH2,Gurnnasson08,Capone10}. The fact that the electron-phonon coupling can play an important part besides the electron-electron interaction takes on particular importance in reduced dimensions where the intrinsically expanded range of electronic instabilities can be further  broadened by  the  interplay between both interactions.  \cite{Hirsch83b,Mazumdar83,Bakrim14} 
 
 In this matter one-dimensional (1D) models of interacting  electrons coupled to a bosonic phonon field have been long  considered  as  primary  models  to study the competing effects of  Coulomb and retarded interactions on  ordered phases at zero temperature. 
In the present work we shall be concerned with two models at half-filling  that have been the focus of considerable attention in the past few decades, namely the 1D Holstein-Hubbard model (HH)\cite{Holstein59,Hirsch85} and   the Su-Schrieffer-Heeger-Hubbard model\cite{Su79}  and its variant for optical phonons called for short the Peierls-Hubbard (PH) model\cite{Hirsch83b,Mazumdar83,Sengupta03}.

The corresponding Hamiltonians  can be introduced through the following common form in  Fourier space,
\begin{align}
 {H} = &  \sum_{k,\sigma} \varepsilon(k) c^{\dag}_{k\sigma}c_{k\sigma} +\sum_{q}  \omega_{q}\Big(b^{\dag}_{q}b_{q} + {1\over 2}\Big)
\cr
 &+  {1\over \sqrt{L}}\sum_{k,q,\sigma}g(k,q)(b^{\dag}_{q}+b_{-q}) c^\dagger_{k+q,\sigma}c_{k\sigma} \cr
 &+   {U\over L} \sum_{q}n_{q\uparrow}n_{-q\downarrow} + {1\over L} \sum_q V_q\, n_q n_{-q}
\label{hamiltonien},
\end{align}
 where $c^{\dag}_{k\sigma}$($c_{k\sigma}$) creates (annihilates) an electron of wave vector $k$ and  spin $\sigma$, and $L$ is the number of lattice sites (the lattice distance $a=1$). The  electron spectrum is 
\begin{equation}
\label{ }
 \varepsilon(k) = -2t \cos k,
\end{equation}
  where $t$ is the hopping amplitude. At half-filling the band is filled up to the Fermi points $\pm k_F  = \pm \pi/2$.
 
In the case of the HH model,   momentum independent intramolecular phonons of energy  $\omega_q\equiv\omega_0$ ($\hbar =1$), described  by the   creation (annihilation)  phonon operators  $b^{\dag}_{q}$($b_{q}$)  of wave vector $q$,   are coupled to electrons through the momentum independent coupling constant 
 \begin{equation}
g(k,q)\equiv {g}_0,\ \ \ \ \ \ \ \ \ \ ({\rm HH})
\end{equation}
where ${g}_0>0$. 

For the Su-Schrieffer-Heeger model,  the phonons belong to an acoustic branch of frequency, 
 \begin{equation}
\omega_q=\omega_D|\sin q/2|,
\end{equation}
 where   $\omega_D$ corresponds to the Debye  frequency for the zone boundary $q=\pi$ phonon, namely at twice the Fermi wave vector $2k_F$ for a half-filled band. Their coupling to electrons results from the modulation of the electron transfer integral by   lattice vibrations  which   leads to the momentum-dependent coupling constant   
\begin{equation}
\label{LambdaD}
g(k,q) =  i{{g}_D} \sin{q\over 2} \cos\Big(k +{q\over 2}\Big) 
\end{equation}
where ${g}_D$ is a constant proportional to the spatial derivative of the hopping integral $t$. Retaining only the coupling to $q=2k_F$ phonons  at the zone edge corresponds to the PH limit. 
 
The Coulomb part of the Hamiltonian (\ref{hamiltonien}) is common to both models. It comprises the Hubbard and  extended-Hubbard interactions $U$ and $V$ ($V_q=V\cos q$) between electrons on each site  and nearest-neighbor sites. These are expressed in terms of the  occupation number, $n_q= n_{q\uparrow}+ n_{q\downarrow}$,  where $n_{q\sigma} = \sum_k c^\dagger_{k+q,\sigma} c_{k\sigma}$. 

The phase diagrams of both models have been the object of  sustained interest during the last decades. To start with the half-filled HH model in the pure Hubbard limit at $V=0$, many efforts have been  devoted to clarifying  its structure. Special emphasis  has been put  on the frontier separating the fully gapped (spin and charge)  $2k_F$ site-centered charge-density-wave (CDW) and the charge (Mott) gapped $2k_F$ spin-density-wave (M-SDW) phases, namely when the amplitude of the phonon-mediated electron coupling  
$$
g_{\rm ph}= 2  {g}^2_0/\omega_0, \ \ \ \ \ {\rm (HH)}
$$  
becomes of the order of the local repulsion $U$. 

 The existence of an intermediate metallic state was  predicted to occur long ago on the basis of real-space renormalization group arguments \cite{Guinea83}, but found to be absent in former world line Monte Carlo simulations\cite{Hirsch85} and two-cut off renormalization group analysis\cite{Caron84,Bindloss05}.  The presence of such a state  was more recently supported  by density matrix renormalization group  (DMRG) and  variational  methods\cite{Jeckelmann99,Takada03}. Its origin was thereafter  extensively discussed from various approaches\cite{Clay05,Tezuka05,Hardikar07,Tam07b,Fehske08,Payeur11}. 
In particular, quantum Monte Carlo (QMC)\cite{Clay05,Hardikar07,Hohenadler13} and further  DMRG \cite{Tezuka07}   calculations have  corroborated its existence over a definite,   $\omega_0-$dependent region of the ($U , g_{\rm ph} ) $ plane. On more  analytical grounds,     the phase diagram of the HH model has been investigated    by the functional RG (fRG) method  at the one-loop level.\cite{Tam07b}  The   frequency dependence of the electron-electron vertices has thus been obtained in  the putative intermediate region.   umklapp scattering, though irrelevant in the static limit,  was found to be  large  at finite frequency. This   was interpreted  as the driving force of a CDW phase gapped in the charge sector, going  against the existence of  an intermediate metallic state, or at the very least the existence of   dominant superconducting correlations  whose  presence in numerical simulations was    attributed to finite-size effects \cite{Clay05,Tam11}. 

In the first part of this paper, we reexamine this problem using a RG method\cite{Bakrim07} similar to the fRG and proceed to a detailed scan   of the phase diagram of the HH model as extracted from  the  singularities of  susceptibilities and the frequency dependence of electron-electron vertices.  We thus reproduce to a high degree of accuracy the   boundaries of the intermediate phase   found by  the QMC simulations of Hardikar {\it et al.,}\cite{Hardikar07} at $g_{\rm ph} \gtrsim U$ and arbitrary $\omega_0$\cite{NoteHH1}. In this sector of the phase diagram, the CDW and to a lesser degree  singlet superconducting (SS) response functions are  found to be singular. This is compatible    with a metallic state  characterized by   effective attractive couplings, which evolves toward those of the attractive Hubbard model  at half-filling when the non adiabatic,      $\omega_0\to \infty$  limit is taken at $U=0$. The present RG calculations   further predict  the presence of a particular structure  of the  boundary between the metallic phase and the M-SDW state. A gapless Luttinger liquid and a gapped $2k_F$  bond-centered charge-density-wave   (BOW) phases are successively found  to enfold the boundary as a function of $U$. Both prevent a  direct passage from  M-SDW to   CDW in the  ($U , g_{\rm ph} )-$plane at  finite $\omega_0$. 

As to the phase diagram of the  PH model at zero temperature, it is known to display a simpler structure.  Numerical simulations of the phase diagram from QMC\cite{Sengupta03} or  DMRG\cite{Pearson11}  techniques for instance  agree  with a direct transition line between the fully gapped  Peierls BOW   (P-BOW) state  and the M-SDW phase in  the $(U, g_{\rm ph})-$plane for small  $V\ge 0$, a feature  of PH phase diagram also well  depicted analytically by the the so-called two-cutoff RG calculations \cite{Caron84,Zimanyi88,Bindloss05}, provided the couplings and Debye  frequency $\omega_D$ are not too large.  The RG calculations  to be   developed  here  in the continuum limit     agree  with QMC results carried out for non dispersive phonons  in  the weak coupling sector for $U$ and $V$.\cite{Sengupta03} 

In Sec.~II,   one-loop flow equations for the scattering amplitudes, self-energy and response functions are given for both models in the electron gas continuum limit. In Sec. ~III, the   flow equations are solved  for the HH and PH models, respectively. The singularities in the couplings at zero and finite frequency, along with static susceptibilities are tracked in the ($U,g_{\rm ph})-$plane and serve to map out the phase diagrams for different $\omega_{0,D}$ and $V$ in the case of the PH model. Comparison with numerical results is made. We conclude in Sec.~IV.

\section{Renormalization group equations of the continuum  limit}
\label{RG}
From the functional integral formulation of the partition function, the  integration  of the phonon field leads to a  (Matsubara) frequency dependent interaction between electrons\cite{Bakrim07}. When   combined with the Coulomb terms $U$ and $V$, this yields  an effective electron-electron interaction of the form:
\begin{equation}
g_{i=1,2,3}( {\omega}_1, {\omega}_2,{\omega}_3)  =  g_i+\frac{g_{{\rm ph},i}}{1 + (\omega_{1}-\omega_{3})^2/ \omega_{0,D}^2}, 
\label{effective_coupling}
\end{equation}
when expressed in the $g$-ology picture of the electron gas model,  where $\omega_i = (2n_i+1)\pi T$. In this continuum framework, the electron spectrum (\ref{hamiltonien}) is  linearized  around the Fermi points $pk_F=\pm k_F$ and becomes 
\begin{equation}
\label{ }
\epsilon(k)\approx\epsilon_p(k) =v_F(pk-k_F).
\end{equation} 
 The spectrum is bounded by   the band width cutoff,   $E_0/2=\pi v_F/2$, on either side of the Fermi level, where $v_F=2t$ is the Fermi velocity at half-filling. The interaction defined on the Fermi points then breaks into three pieces, namely the backward ($g_1$), forward ($g_2$) and umklapp ($g_3$) scattering amplitudes whose frequencies satisfy $\omega_1+\omega_2=\omega_3+\omega_4$. 

The corresponding attractive amplitudes for the phonon-mediated part are given by 
\begin{equation}
\label{HH}
g_{{\rm ph},1,2,3}=  -g_{\rm ph} , \ \ \ \ \ {\rm (HH)}
\end{equation}
for the HH model, where $g_{\rm ph}= {2g^2_0/\omega_0}$.

For the  PH model, however, the electron phonon-matrix element is momentum dependent, which yields
\begin{equation}
\label{PH}
g_{{\rm ph},1}= -g_{{\rm ph},3}=   -g_{\rm ph} ;\ \ \ \  g_{{\rm ph},2}=0,  \ \ \ \ \ {\rm (PH).}
\end{equation}
 where 
 $$ 
  g_{\rm ph}= {2g^2_D/ \omega_D}. \ \ \ \ \ {\rm (PH)}
 $$   Therefore   repulsive umklapp and the absence of  forward scattering at the bare level are direct consequences of the momentum dependence of the bond  electron-phonon interaction (\ref{LambdaD}). This is a key ingredient for the differences between HH and  PH models. 

For the  Coulomb part   that is common to both models, we have the familiar bare coupling combinations of  the electron gas  at half-filling,\cite{Emery79,Solyom79}
\begin{equation}
\label{UV}
g_1=g_3=U-2V, \ \ \ \ g_2=U+2V.
\end{equation}
All the bare   amplitudes are confined to the weak-coupling regions $\tilde{g}_i\equiv g_i/\pi v_F <1$ and $\tilde{g}_{{\rm ph} }\equiv {g}_{{\rm ph}}/\pi v_F <1$.

We apply a   Kadanoff-Wilson  RG procedure for the partition function,\cite{Bourbon91,Bakrim07} which consists at zero temperature of the successive integration of electronic momentum degrees of freedom from the cutoff energy $\pm E_0/2$ down to the energy distance $\pm E_0(\ell)/2$ on either side of the Fermi level. Here   $E_0(\ell)= E_0 e^{-\ell}$ is  the effective bandwidth at step $\ell$. This successive integration in the momentum is performed for all Matsubara frequencies. The   frequency axis 
is   divided  into  61 sections or patches between the cutoff 
values $\pm \omega_{\rm max}=\pm 1.5E_0/2$, which serve as bounds of integration
for the frequency. The interactions are taken   constant over
each patch  where the loop integrals are done exactly. The successive integration leads to  the flow of coupling constants, one-particle self-energy,  and susceptibilities. At the one-loop level, the flow of the normalized scattering amplitudes at $T\to 0$  reads
 \begin{widetext}
\begin{eqnarray}\label{g1}
\partial_\ell \tilde{g}_{1}\left(\omega_{1},\omega_{2},\omega_{3}\right)& = & 2\int_{\omega'} \ \Big\{\big[-\tilde{g}_{1}(\omega_{1},\omega',\omega_{3})\tilde{g}_{1}(\omega_{4},\omega',\omega_{2})-\tilde{g}_{3}(\omega_{1},\omega',\omega_{3})\tilde{g}_{3}(\omega_{4},\omega',\omega_{2})+\tilde{g}_{1}(\omega_{1},\omega',\omega_{3})\tilde{g}_{2}(\omega',\omega_{4},\omega_{2})
\nonumber\\
&+ &   \ \tilde{g}_{3}(\omega_{1},\omega',\omega_{3})\tilde{g}_{3}(\omega',\omega_{4},\omega_{2})\big] I_{+}(\omega',\omega_{1}-\omega_{3})-\tilde{g}_{1}(\omega_{1},\omega_{2},\omega')\tilde{g}_{2}(\omega_{3},\omega_{4},\omega') I_{-}(\omega',\omega_{1}+\omega_{2})\Big\},\\
\label{g2}
\partial_\ell \tilde{g}_{2}\left(\omega_{1},\omega_{2},\omega_{3}\right)&=&\int_{\omega'}\Big\{\big[-\tilde{g}_{1}(\omega_{1},\omega_{2},\omega')\tilde{g}_{1}(\omega_{3},\omega_{4},\omega')-\tilde{g}_{2}(\omega_{1},\omega_{2},\omega')\tilde{g}_{2}(\omega_{3},\omega_{4},\omega')] I_{-}(\omega',\omega_{1}+\omega_{2})
\nonumber\\
&&\ \ \ \ \   +\ \ \big[ \, \tilde{g}_{2}(\omega',\omega_{2},\omega_{3})\tilde{g}_{2}(\omega',\omega_{4},\omega_{1})
+ \, \tilde{g}_{3}(\omega',\omega_{2},\omega_{3})\tilde{g}_{3}(\omega',\omega_{4},\omega_{1})] I_{+}(\omega',\omega_{2}-\omega_{3})\Big\},
\\
\label{g3}
\partial_\ell \tilde{g}_{3}\left(\omega_{1},\omega_{2},\omega_{3}\right)&=&2\int_{\omega'}\Big\{\big[-2\tilde{g}_{1}(\omega_{1},\omega',\omega_{3})\tilde{g}_{3}(\omega_{4},\omega',\omega_{2})+\tilde{g}_{1}(\omega_{1},\omega',\omega_{3})\tilde{g}_{3}(\omega',\omega_{4},\omega_{2})
\nonumber\\
&&\ \ \  +\tilde{g}_{2}(\omega',\omega_{4},\omega_{2})\tilde{g}_{3}(\omega_{1},\omega',\omega_{3})] I_{+}(\omega',\omega_{1}-\omega_{3}) +\tilde{g}_{2}(\omega',\omega_{2},\omega_{3})\tilde{g}_{3}(\omega',\omega_{4},\omega_{1})
 I_{+}(\omega',\omega_{2}-\omega_{3})\Big\},
 \end{eqnarray}
\end{widetext}
where $\partial_{\ell} \equiv \partial/\partial \ell$ and $\int_{\omega'} \equiv \int^{+\omega_{\rm max}}_{-\omega_{\rm max}} d\omega'/ 2\pi$. Here $I_\mp$ refer  to the $2k_F$ electron-hole (Peierls) and electron-electron (Cooper) loops,
\begin{eqnarray}
&& I_{\mp}(\omega,\Omega)   =    \cr
&&  \Lambda_\ell  \frac{(\omega-\Sigma (\omega))(\omega\mp\Omega\pm\Sigma (\Omega\pm\omega))  +\Lambda_\ell^2}{[(\omega-\Sigma (\omega))^2  + \Lambda_\ell^2][(\omega\mp\Omega\pm\Sigma (\Omega\pm\omega))^2 +  \Lambda_\ell^2]}, \nonumber\\
\label{I}
\end{eqnarray}
where $\Lambda_\ell=E_0(\ell)/2$. The imaginary part of the Matsubara self-energy,  $\Sigma$, for right or left-going electrons obeys   the equation
\begin{eqnarray}
\partial_\ell\:\Sigma(\omega ) &=  &\int_{\omega'}
  \Big\{   \tilde{g}_{1}(\omega',\omega,\omega_{1})-2\tilde{g}_{2}(\omega,\omega',\omega)\Big\} \cr
&& \ \ \ \  \times \Lambda_{\ell}\frac{\omega'-\Sigma(\omega')}{[\omega'-\Sigma(\omega')]^{2}+\Lambda^{2}_{\ell}},
\end{eqnarray}
with the initial condition $\Sigma(\omega)|_{\ell=0}=0$. The one-loop flow is essentially governed by the superimposition of Cooper and Peierls pairing channels which  interfere with one another, mixing  ladder, closed loop, and vertex correction diagrams at every order.    The combination  controls the nature  of correlation at long distance, and in the presence of a phonon part will be dependent  on the degree of retardation or the phonon frequency $\omega_{0,D}$.

The nature of correlations as a function of $\ell$ are   analyzed from a selected set of static normalized susceptibilities $\tilde{\chi}_\mu (\equiv \pi v_F \chi_\mu)$, namely those likely to be singular in the density-wave and superconducting channels. The static susceptibilities obey an  equation of the form
\begin{eqnarray}
\label{Ki}
\partial_\ell \tilde{\chi}_\mu &=& \int_{\omega}  |z_{\mu}(\omega)|^{2}I_{\mp}(\omega,0). 
\end{eqnarray}
Each of these involves a pair vertex function $z_\mu$ governed by the equation   
\begin{eqnarray}
\label{z}
\partial_\ell \:z_{\mu}(\omega) \!\!&=& \!\!\int_{\omega'}  \tilde{g}_\mu (\omega',\omega,\omega)z_{\mu}(\omega')I_{\mp}(\omega',0),  
\end{eqnarray}
where $z_\mu(\omega)|_{\ell=0}=1$.
For the charge  density-wave channel, we  shall consider  the  CDW   and BOW susceptibilities at the wave vector  $2k_F$, and to which  correspond   the combinations of couplings
\begin{align}
\label{gBOW}
    \tilde{g}_{\mu}(\omega',\omega,\omega)\Big|_{\mu= {\rm CDW(BOW)}} & =    \tilde{g}_{2}(\omega',\omega,\omega)-2 \tilde{g}_{1}(\omega,\omega',\omega) \cr
  &  \pm\  [\tilde{g}_{3}(\omega,\omega',\omega')\!-\!2\tilde{g}_{3}(\omega',\omega,\omega')].
\end{align}
For the spin-density-wave channel,  the antiferromagnetic  or site centered  SDW  susceptibility may  be also singular; it  corresponds to the combination of couplings
\begin{eqnarray}
   \tilde{g}_{\mu={\rm SDW}}(\omega',\omega,\omega) & =   & \tilde{g}_{2}(\omega',\omega,\omega)\\
  &- &\!\!  [\tilde{g}_{3}(\omega,\omega',\omega')\!-\!2\tilde{g}_{3}(\omega',\omega,\omega')].\nonumber
  \label{gSDW}
\end{eqnarray}
 In the superconducting channel, only the SS susceptibility may develop a singularity. It is linked  to the combination, 
\begin{eqnarray}
\tilde{g}_{\rm SS}(\omega',\omega,\omega) =  -\tilde{g}_{2}(\omega,-\omega,\omega')- \tilde{g}_{1}(-\omega,\omega,\omega').
\end{eqnarray}

The solution of equations as a function of $\ell$ for $\chi_\mu$ and the scattering amplitudes reveals the singularities in $\chi_\mu$; these allow the identification of dominant and subdominant singular correlations\cite{Bakrim07}. A divergence at a finite $\ell_c$ indicates the breakdown of the weak-coupling one-loop RG procedure. Nevertheless, it singles out an energy scale or a gap $\Delta =E_Fe^{-\ell_c}$ at zero temperature that correlates with   a gap in the spin or/and charge degrees of freedom at $\ell_c$. The gap in the spin sector is associated with a singularity in   the static, $\omega \to 0$,  limit for the attractive backward-scattering amplitude (\ref{g1}), whereas in the charge sector,  a gap is found for   singular either static repulsive or attractive  umklapp  (\ref{g3}).
\section{Results and Discussion}
\subsection{ Holstein-Hubbard model}
\begin{figure}
 \includegraphics[height=11cm,width=9.0cm]{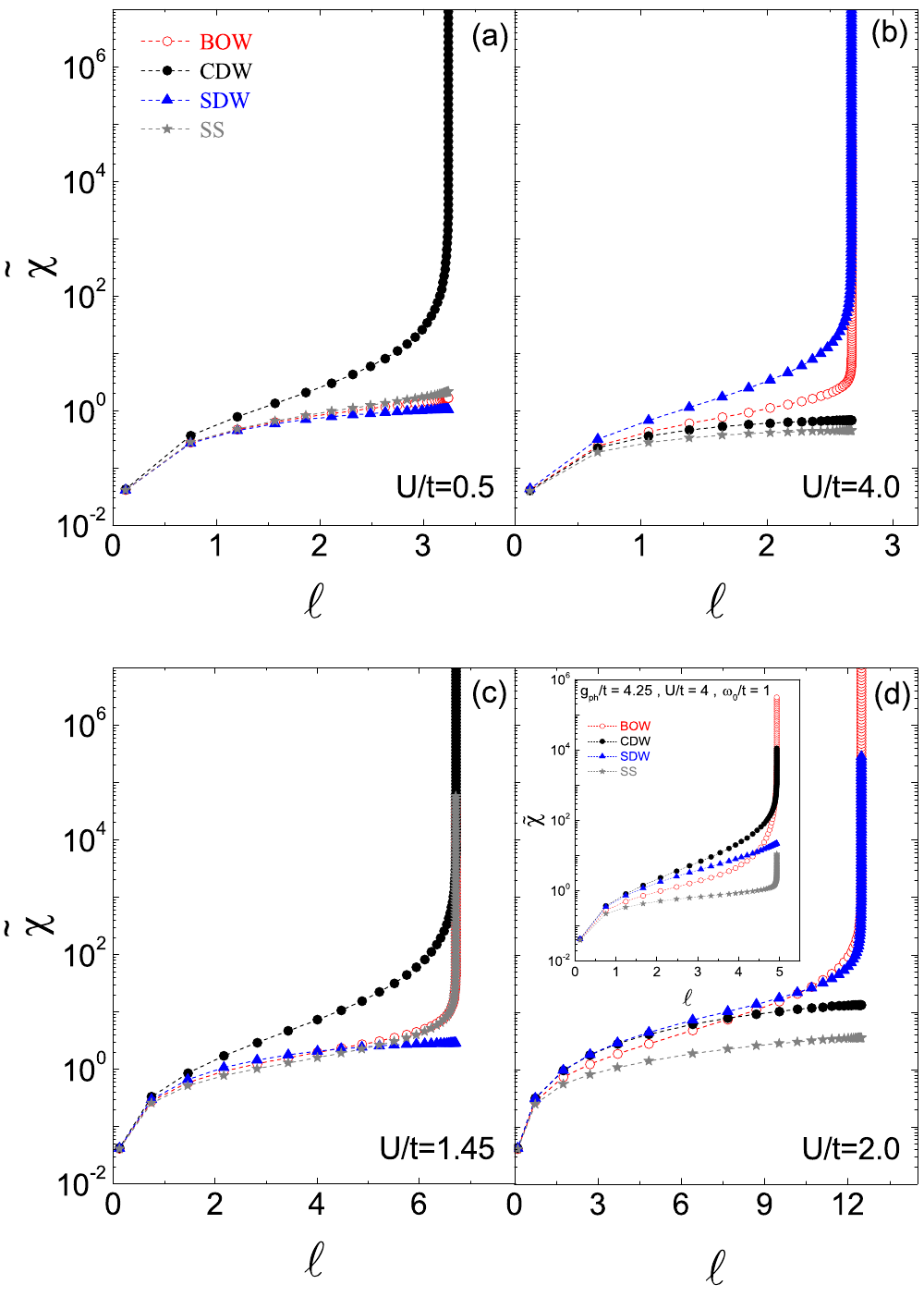}
\caption{\label{fig1} (Color online) In (a)-(c), the flow  of normalized susceptibilities $\tilde{\chi}_\mu$  {\it vs} $\ell$ for the Holstein-Hubbard model    for $g_{\rm ph }/t= 2$ at different values of $U/t$. In (d), the flow of susceptibilities on the boundary of spin-density-wave phase with the intermediate phase;  insert: susceptibilities on the boundary with the charge-density-wave phase.    In all cases the     phonon frequency is fixed  at $\omega_{0}=t$.}
\end{figure}
 
Let us first consider the HH model at $U>0$. In Fig.~\ref{fig1}  typical $\ell$ dependencies for the susceptibilities are shown  for   a  fixed bare  phonon-mediated   coupling amplitude,  $g_{\rm ph}  =2t$,  and phonon frequency  $\omega_0=t$.
At  $U=0.5 t$, that is in the region well above the  bisecting line $  g_{\rm ph}=U$, a   divergence is found  only for the CDW susceptibility [Fig.~\ref{fig1} (a)]. 
  The value of $\ell_c\simeq 3.25$ extracted from the Fig.~\ref{fig1} (a) leads to a one-loop CDW gap $\Delta=E_F e^{-\ell_c}$ that is sizably reduced compared to  the adiabatic  --mean field--  value \cite{Bakrim07} $\Delta_0=E_Fe^{-1/(2\tilde{g}_{\rm ph})}$, of the pure phonon model  at $\omega_0\to0$  and for which, $\ell_0 =\pi/2$  at the  coupling considered here. The value of $\Delta$     is also smaller but  closer   to the non adiabatic  $\omega_0\to \infty$  limit of Eqs.~(\ref{g1}-\ref{g3}) at $U=0$,   corresponding to an effective attractive Hubbard model, where $\Delta_\infty = E_F e^{-1/\tilde{g}_{\rm ph}}$  and $\ell_\infty = \pi<\ell_c$ at the one-loop level \cite{Emery79}.    Repulsive $U$, which   primarily reduces double occupancy of electrons on sites decreases   CDW correlations and  in turn their coupling to $2k_F$ molecular phonons through vertex corrections in Eqs. (\ref{g1})--(\ref{g3}), can explain these downward shifts of the gap   \cite{Caron84,Bindloss05}. 
  \begin{figure}
 \includegraphics[height=10cm,width=8.0cm]{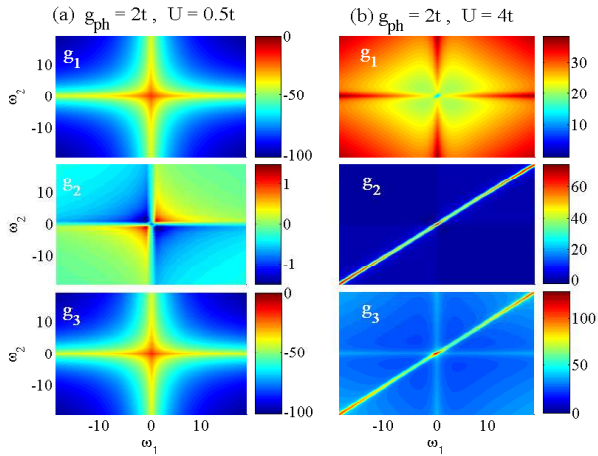}
\caption{\label{fig2} (Color online) Contour plot of the frequency dependent normalized couplings $\tilde{g}_{i}(\omega_{1},\omega_{2},\omega_{1})$  in a finite portion of the $(\omega_1,\omega_2)$ plane   at $g_{\rm ph}=2t$, $\omega_{0}=t$,  $\ell$ close to the critical  $\ell_c$, and  for two values of Coulomb interaction:  (a) $U=0.5t$ (CDW)  and (b)   $U=4t $ (M-SDW).  }
\end{figure}
  
  Regarding now the structure of scattering amplitudes at finite frequency in this CDW phase, we follow  Ref.~\cite{Tam07b} and show in Fig.~\ref{fig2} (a) the contour plot of the projected   $\tilde{g}_{i=1,2,3}(\omega_1,\omega_2,\omega_3)$   in the $(\omega_1,\omega_2)$-plane when $\ell \approx \ell_c$. As  previously noted in the framework of fRG\cite{Tam07b}, the scattering amplitudes display significant structure as a function of frequency, notably an attractive  singular behavior  for the CDW phase that persists away from the origin of the $(\omega_1,\omega_2)$-plane. Divergences  toward negative values are   seen   in both $g_1(\omega_1,\omega_2,\omega_1)$ and $g_3(\omega_1,\omega_2,\omega_1)$  with maxima at the origin [Fig.~\ref{fig2} (a)],  supporting the existence of a gap in both spin and charge degrees of freedom  compatible with the absence of enhancement in spin correlations and  the existence of commensurate CDW order at half-filling.
  
  One can wonder about the impact of vertex    singularities in the finite frequency range on correlations. In the present    RG procedure  both the marginal and    irrelevant (frequency) parts of couplings are comprised in the   three-variables scattering amplitudes  $g_i(\omega_1,\omega_2,\omega_3)$. However, one can get a rough idea  of how the frequency dependence emerges.  Assuming that the amplitudes can be expanded around the origin as
  
  $$
  g_i(\omega_1,\omega_2,\omega_3) = g_i(0,0,0) + \sum_{j=1}^3 \bar{g}_{i,j} \omega_j^2 + \ldots, 
  $$ 
we see   that the amplitudes, $\bar{g}_{i,j}$, etc.,  of the frequency dependent part  are by power counting superficially irrelevant in the RG sense. However, these will be  coupled to the marginally relevant $g_i(0,0,0)$ at the one-loop level and may become in turn relevant when one or more $g_i(0,0,0)$ scales to strong coupling, that is   as $\ell$ approaches $\ell_c$. This is likely what happens in the present RG procedure.   
Being strongly weakened  by the frequency convolution of  couplings and  single loops (Peierls and Cooper) which decay  as $\omega^{-2}$ [Eqs.~(\ref{I}) and (\ref{z})], these high frequency singularities have in the end a limited  impact on susceptibilities and then correlations;  the corrections remaining in most cases at  a quantitative level. In special regions of the phase diagram,  however, the  consequence  can be more substantial and can act on the nature of the ground state as  will be discussed below.

 If we now move   on the opposite side of the  bisecting line, on the Hubbard side of the phase diagram at $U=4t$,   the strongest singularity is found for the SDW susceptibility at $\ell_c\simeq 2.7$, which comes with  subdominant singular   BOW correlations [Fig.~\ref{fig1} (b)]. These are the  characteristics of the repulsive Hubbard model at half-filling which presents  a charge or Mott gap $\Delta$  at $\ell_c$, in accord with   singular repulsive $g_2(0,0,0)$ and umklapp $g_3(0,0,0)$  scatterings at zero frequency as meant in Fig.~\ref{fig2} (b). As for the backscattering, the same Figure shows that it  becomes repulsive and large close to $\ell_c$ over the whole frequency range, with  only a minimum for $g_1(0,0,0)$ [$\ll g_{2,3}(0,0,0)$].  Nevertheless,   relevant repulsive backscattering introduced by the dynamics leaves the spin sector gapless as shown by the singular behavior of the SDW response in  Fig.~\ref{fig1} (b). One also notes that the strong repulsive coupling  that characterizes the backscattering amplitude $g_1(\omega_1,\omega_2,\omega_3)$ at finite frequency  has a limited influence on correlations,  reducing  for example the amplitude of BOW correlations compared to the situation of the pure Hubbard model.

 Regarding the amplitude of the gap, the   finite attractive contribution coming  from phonons [Eq.~(\ref{HH})] weakens  the amplitude of   $\Delta$ compared to the pure   Hubbard result \cite{Emery79}, $\Delta_\rho = E_F e^{-1/\tilde{U}}$, which  yields $\ell_\rho= \pi/2< \ell_c $, at $U=4t$.   
\subsubsection{Intermbehaviorediate phase}
\begin{figure}
 \includegraphics[height=9cm,width=8.0cm]{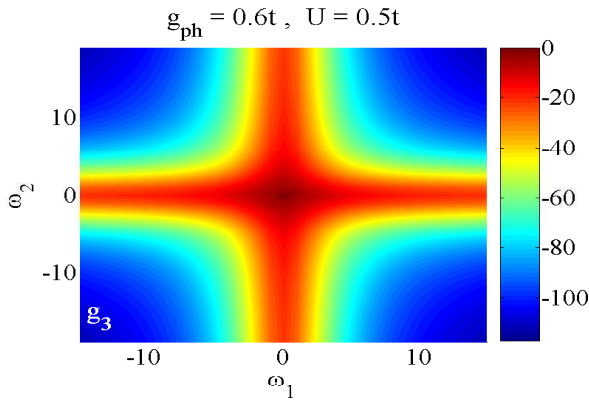}
\caption{\label{fig3} (Color online) Typical contour plot of umklapp scattering amplitude $g_3(\omega_1,\omega_2,\omega_1)$ in the $(\omega_1,\omega_2)$-plane close to $\ell_c$ in   the intermediate I-phase at $\omega_0=t$.}
\end{figure}

As a function of $U$, the ground state  evolves from the fully gapped CDW  to the M-SDW states   discussed above.  Within the CDW state the reduction of the gap $\Delta$ with $U$ carries on up to a critical value where   static umklapp, $g_3(0,0,0)$, is no longer singular, but goes to zero  and  becomes irrelevant (Fig.~\ref{fig3})\cite{Hirsch85,Tam07b}, whereas $g_1(0,0,0)$ and $g_2(0,0,0)$ remain attractive as in Fig.~\ref{fig2} (a). The irrelevance of the static $g_3$ is concomitant with the onset  of an  additional, but subdominant singular susceptibility in the SS channel, as shown in Fig.~\ref{fig1} (c). From the same figure, we also note   that the BOW response is also  enhanced.   The CDW-SS singular correlations  persist   over a finite $U$-interval    before the system ultimately enters in the  M-SDW  phase described above [Fig.~\ref{fig1} (b)]. A detailed scan in the $(U,g_{\rm ph})$-plane allows us to outline  an  entire intermediate (I) region of  the phase diagram of Fig.~\ref{fig2} (b)  where  similar  features for CDW and SS correlations are found. Note that within the I region, we observe a reinforcement of the SS correlations  as $U/t$ decreases; its singular behavior reaching  the CDW one as $U\to 0$ and $\omega_0 \gg t$, which is essentially the situation of the attractive Hubbard model.

\begin{figure}
 \includegraphics[height=11cm,width=7.5cm]{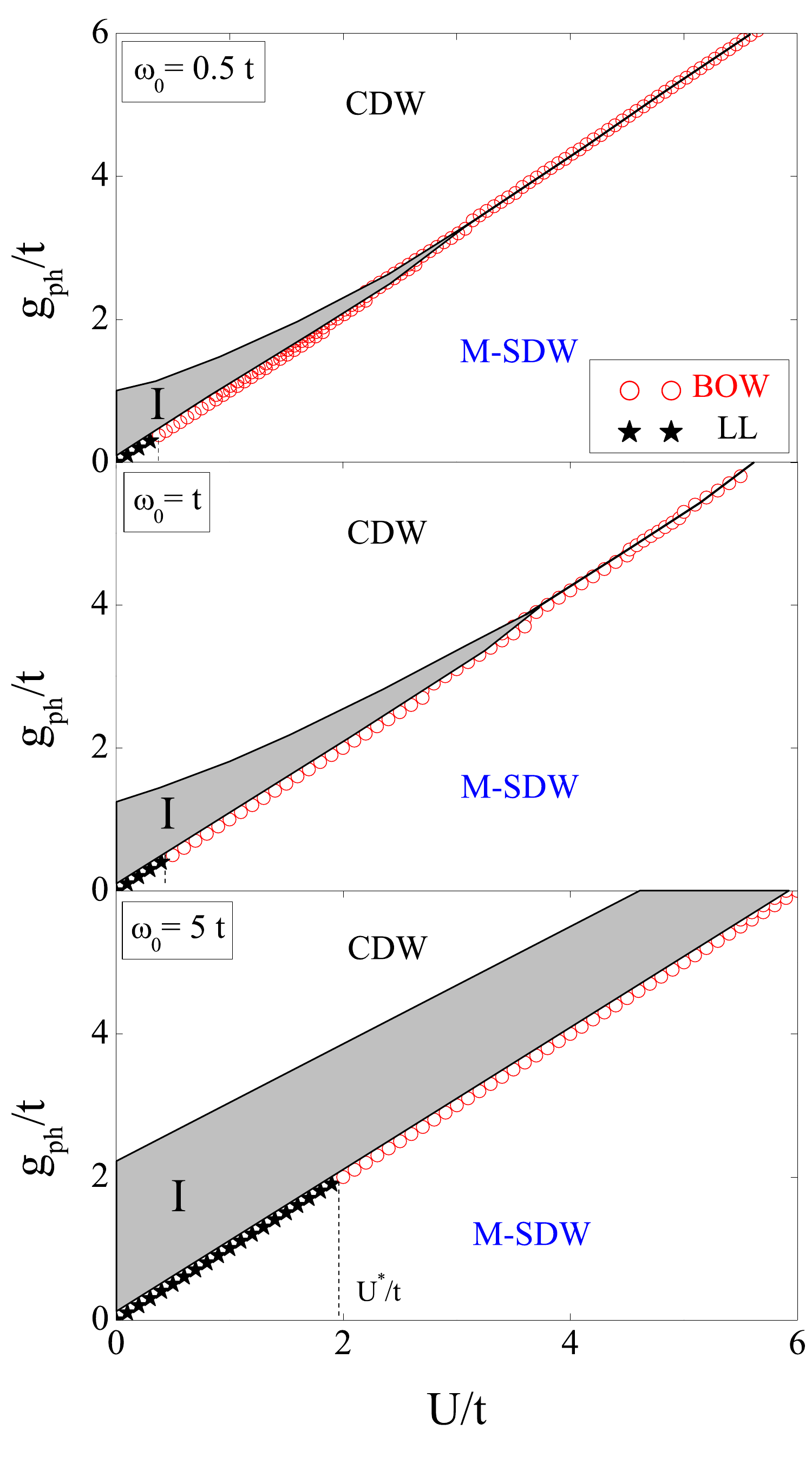}
\caption{\label{fig4} (Color online) One-loop RG  phase diagram   of the   1D Holstein-Hubbard model at   different molecular phonon frequencies:  $\omega_{0}=0.5t$ (a), $\omega_{0}=t$ (b) and $\omega_{0}=5 t$ (c).}
\end{figure}

   The     boundary  of the I state closes at some  point   $(U^c,g^c_{\rm ph})$, whose locus  depends on the phonon frequency  $\omega_0$. The resulting phase diagrams shown in Fig.~\ref{fig4} for different $\omega_0$  stand particularly well the comparison with   previous numerical analysis,  notably those carried out by the    QMC \cite{Hardikar07} and   DMRG \cite{Tezuka05} methods. Comparing for instance the metallic I region obtained  from the QMC simulations of Hardikar and Clay (see Fig.~8 of Ref.~\cite{Hardikar07}) with the one   deduced here by RG, a quite accurate  match is found  for  an amazing  range of interactions and   phonon frequencies.    

As to the  the origin of the I phase, it must be stressed at the outset that within the electron gas model, irrelevant  umklapp and strong attractive backward scattering, along with  degenerate CDW and SC singular susceptibilities,  are well known characteristics of the attractive 1D Hubbard model at  half filling \cite{Emery79,Solyom79,Giamarchi04}.  For the HH model, this limit is clearly realized  at $U=0$,  along the $g_{\rm ph }$ axis   as $\omega_0\to \infty$ (see note in Ref.\cite{NoteHH1}).   As  $\omega_0$  is  increased beyond $t$, Fig.~\ref{fig4} shows indeed an ever-growing  I region which  includes the $(0,g_{\rm ph})$ axis. This  suggests   that the whole region    is   governed     by a Luther-Emery type of  fixed point \cite{Emery79,Voit95}, characterized by effective attractive couplings and a gap in the spin sector, despite finite retardation and   repulsive $U$ \cite{Hohenadler15}.  Both split the degeneracy  between CDW and SS correlations reinforcing the former with respect to the latter.  According to the   combination  of couplings (\ref{gBOW}) for CDW susceptibility, singular attractive umklapp scattering at finite frequency does have an impact, thought limited,  in  favoring an increase of CDW correlations and lack of degeneracy between CDW and SS  in the I region [Fig.~\ref{fig1} (c)]. 

  The above results  are compatible with those   previously found  from  fRG by Tam {\it et al.} \cite{Tam07b}   for selected points   of the phase diagram well outside and inside the supposed I region.  A different view was held, however,  as to the properties of  charge degrees of freedom in this specific part of the phase diagram. The presence  of singular and  negative umklapp scattering at   finite  frequency found in the I  region (Fig.~\ref{fig3}), was interpreted as the driving force of  the  CDW state, whose charge sector was  then considered    still gapped and insulating.  We consider that  a spatially uniform charge gapped state, if it exists,  must be manifest in the equilibrium  --static-- properties  rather than in the dynamics. Otherwise the Mott-insulating  state thus obtained  would be  hard to reconcile  with a finite charge  compressibility,\cite{Hardikar07} and singular superconducting correlations. 
\subsubsection{Structure of the phase  boundary}
\begin{figure}
 \centerline{\includegraphics[height=6cm,width=7.0cm]{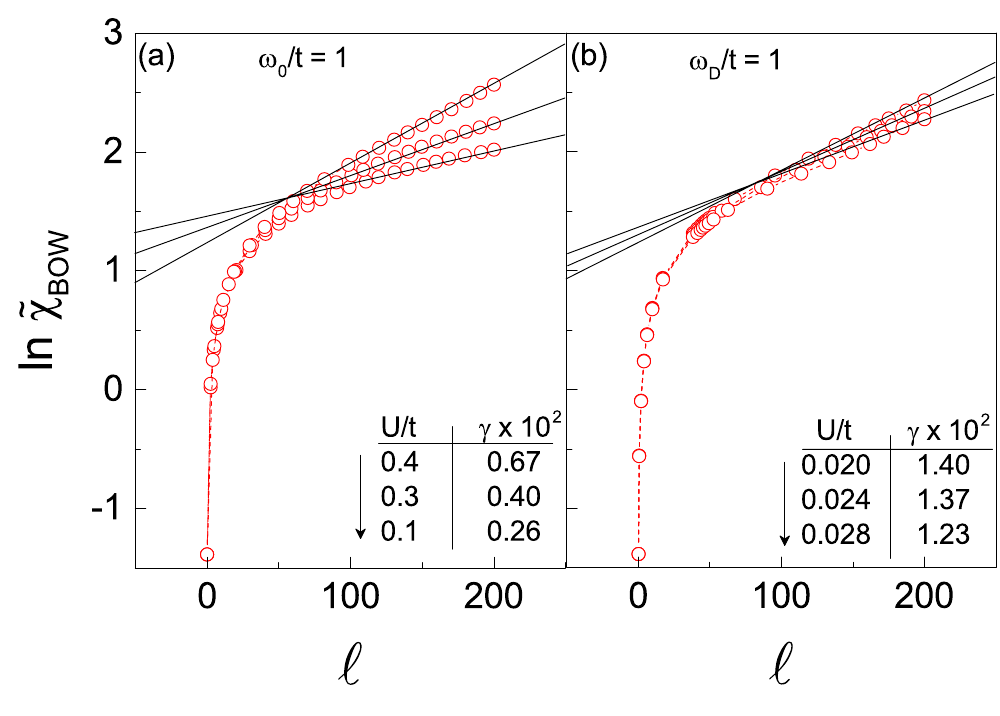}}
\caption{\label{fig5} (Color online) The power-law behavior \hbox{$\chi_{\rm BOW} \sim [E_0(\ell)]^{-\gamma}$} (straight lines)  of the BOW susceptibility on the LL line of the phase diagram at different $U$ for the HH model  in (a) [Fig.~\ref{fig4}~(b)], and for the PH model in (b) [Fig.~\ref{fig9}~(c)]. In both cases, $\omega_0/t=1$.}
\end{figure}

\begin{figure}
 \centerline{\includegraphics[height=10cm,width=7.0cm]{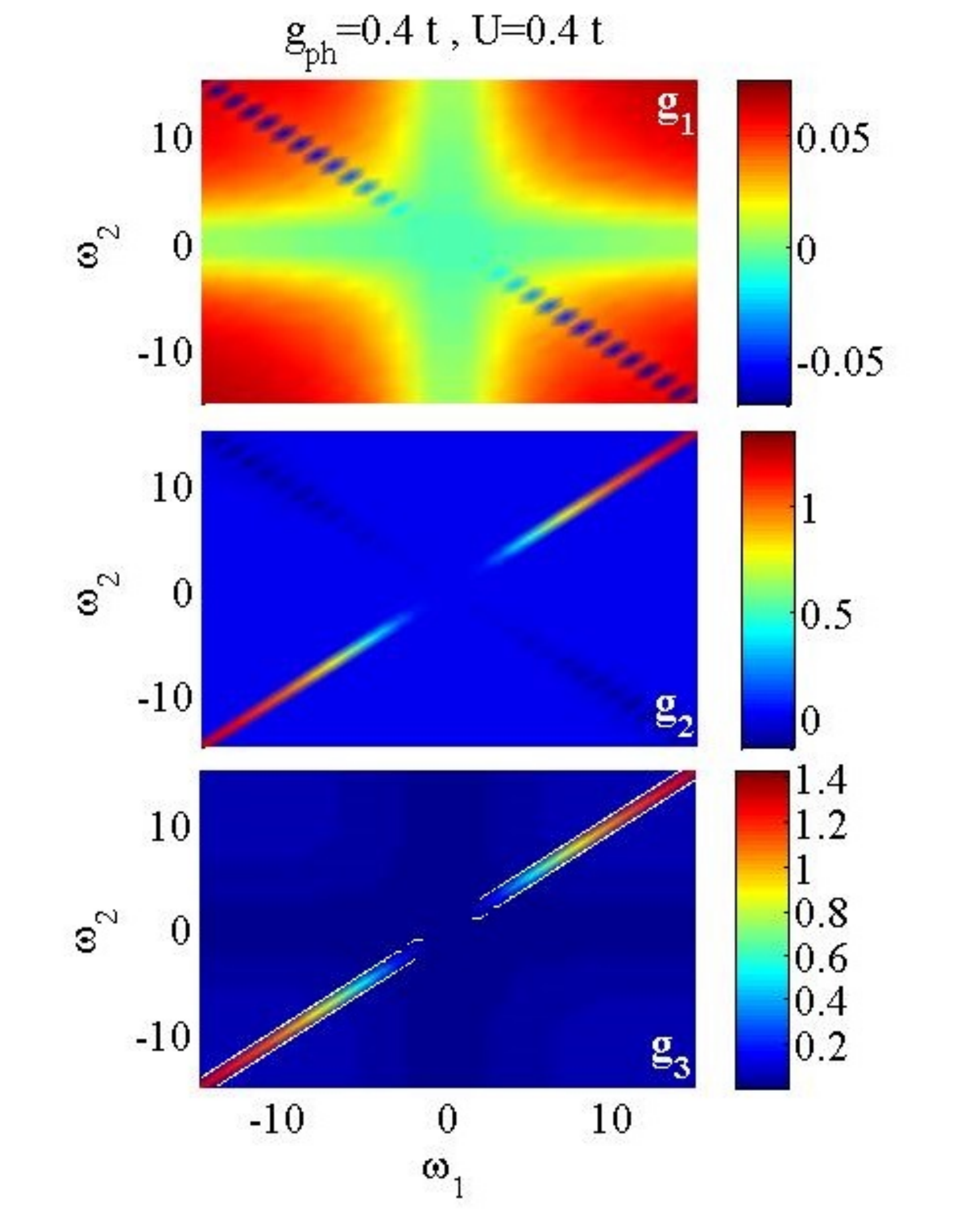}}
\caption{\label{fig6} (Color online) Contour plots of the scattering amplitudes $g_i(\omega_1,\omega_2,\omega_1)$ in the $(\omega_1,\omega_2)$-plane at large $\ell$  and $\omega_0=t$ for the metallic Luttinger liquid (LL) line of the HH model. }
\end{figure}

We close this subsection by a detailed examination of the  boundary to the  M-SDW  state in  the phase diagram.  As one moves along the  bisecting line, $g_{\rm ph}=U$, the RG calculations reveal the existence of a  gapless  transition line  between the M-SDW and  I  phases  (Fig.~\ref{fig4}).   At small $U$ along this line,   no singularity is encountered at   finite $\ell$   in the susceptibilities. These rather exhibit a power law singularity of the form $\tilde{\chi}  \sim  [E_0(\ell)]^{-\gamma}$  at large $\ell$ as shown in Fig.~\ref{fig5}~(a).  The exponent  $\gamma$ is non universal, thought very small and positive,  and     is the largest for the BOW susceptibility.  Within one-loop approximation, such a  power law   is characteristic  of a Luttinger liquid (LL) with effective  very  weak   repulsive interactions at low energy. This is confirmed by the  contour plots of Fig.~\ref{fig6} for the couplings in the ($\omega_1,\omega_2)$ plane, where at large $\ell$ all the couplings are vanishingly small  at the origin and remain weak at finite frequency.  
The LL line terminates on the  bisecting line  at the finite value $U^*=g^*_{\rm ph}$, which increases with $\omega_0$ (Fig.~\ref{fig4}). It has been checked  that in  the limit of large $\omega_0$, the LL and $U=g_{\rm ph}$ lines merge  over the entire  range of couplings with $\gamma\to 0$,  consistently with vanishing initial couplings in (\ref{effective_coupling}).  It is worth noting that the existence of a LL metallic phase with a similar $\omega_0$ dependence, albeit on a  larger area of the phase diagram   has been already pointed out by Fehske {\it et al.}~\cite{Fehske08} using the DMRG technique, thought constrained by possible  finite-size effects.

\begin{figure}
 \centerline{\includegraphics[height=8cm,width=11.0cm]{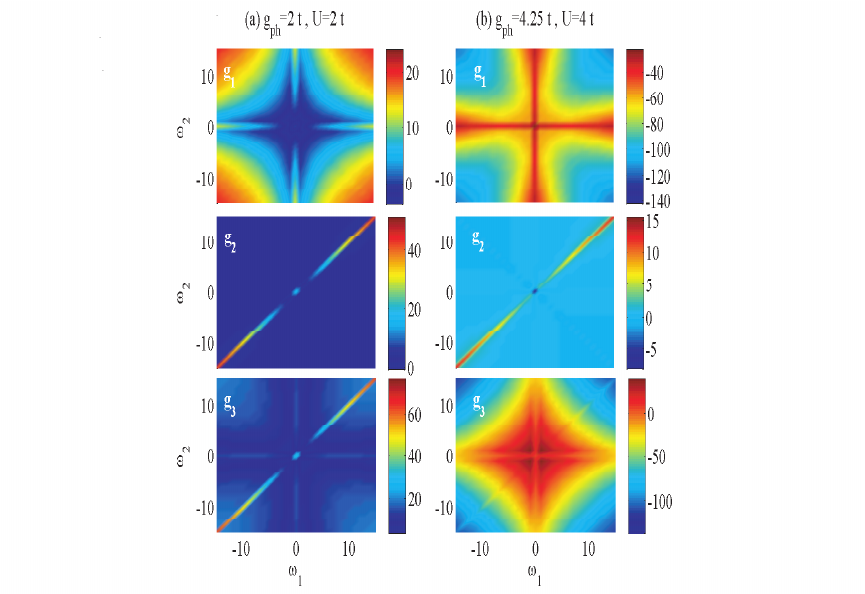}}
\caption{\label{fig7} (Color online) Contour plots of the scattering amplitudes $g_i(\omega_1,\omega_2,\omega_1)$ in the $(\omega_1,\omega_2)$-plane for 	 the HH model  in the BOW regime of the boundary with the intermediate phase at small $U$ (a) and with the CDW phase at larger  $U $ (b). }
\end{figure}

Another  feature   emerges at finite $\omega_0$ where  the ending point $(U^*,g^*_{\rm ph})$ in  Fig.~\ref{fig4}    marks the beginning  of a different boundary with the  M-SDW phase.  For $U> U^*$ a singularity  appears in the susceptibilities at    finite $\ell_c$, the strongest being  for   BOW, whereas  SDW  correlations  become  subdominant  [Fig.~\ref{fig1}~(d)]; no   enhancement in CDW correlations is found.  Actually, as one hits  the    boundary from below the relative importance of SDW and BOW  correlations is inverted compared to the   M-SDW phase [see Figs.~\ref{fig1}~(b), (d)].   The frequency profile of coupling constants in this BOW phase along the boundary is of interest.   For small $g_{\rm ph}/t$ and finite $\omega_0$, as shown in Fig.~\ref{fig7}~(a),  both $g_2(0,0,0)$ and $g_3(0,0,0)$ scale to strong repulsive coupling as $\ell \to \ell_c$, which is indicative of a charge gap. However, $g_1(\omega_1,\omega_2,\omega_1)$ remains relatively small, although  {\it attractive} at   zero frequency and its close vicinity.  It is from these frequency effects, a consequence of retardation, that comes the  origin of dominant BOW correlations.   According to the combination of couplings for the BOW susceptibility in Eq.~(\ref{gBOW}),  a change of sign in the backscattering  in the  low frequency range is sufficient to reinforce BOW  correlations against SDW [Eq.~(\ref{gSDW})].  

 As $g_{\rm ph}/t$ and $U$ become  larger at finite $\omega_0$ along the boundary, $g_1(\omega_1,\omega_2,\omega_1)$ develops  much stronger  attraction at the origin and beyond, suggesting the presence of a spin gap at $\ell_c$. This is reflected by  the absence of singular  SDW correlations in Fig.~\ref{fig1}~(d) (insert). While $g_2(0,0,0)$ becomes small as shown in Fig.~\ref{fig7}~(b), $g_3$ remains positive at  the origin and its neighbourhood, which indicates that at sizable $g_{\rm ph}$ the BOW order is now gapped in the spin sector and  prevails over  CDW that  also becomes singular on the boundary. By cranking up further  $g_{\rm ph}/t$ at finite $\omega_0$ along the boundary, namely beyond the limitations of the RG, it is likely that BOW would be suppressed and the system  would evolve  toward a CDW  ground state.   
The  BOW phase emerging  in this part of the CDW--M--SDW  boundary, as a consequence of retardation at finite $\omega_0$, is reminiscent of  the repulsive 1D extended Hubbard lattice model where the BOW phase is known to  enfold  the \hbox{$U=2V$} line~\cite{Sandvik04}. This separatrix  is known to separate  the same CDW and M-SDW ground states in the continuum approximation \cite{Emery79}.  The 1D extended Hubbard model being  defined on a lattice, however,    it is  the  {\it  momentum dependence} of couplings, though  irrelevant in the RG sense,  that    break   the CDW$-$M-SDW  degeneracy in favor of a BOW state \cite{Tam06,Menard11}. 
\subsection{Peierls-Hubbard model}
\subsubsection{V=0 case}
 \begin{figure}
 \includegraphics[height=6cm,width=8.0cm]{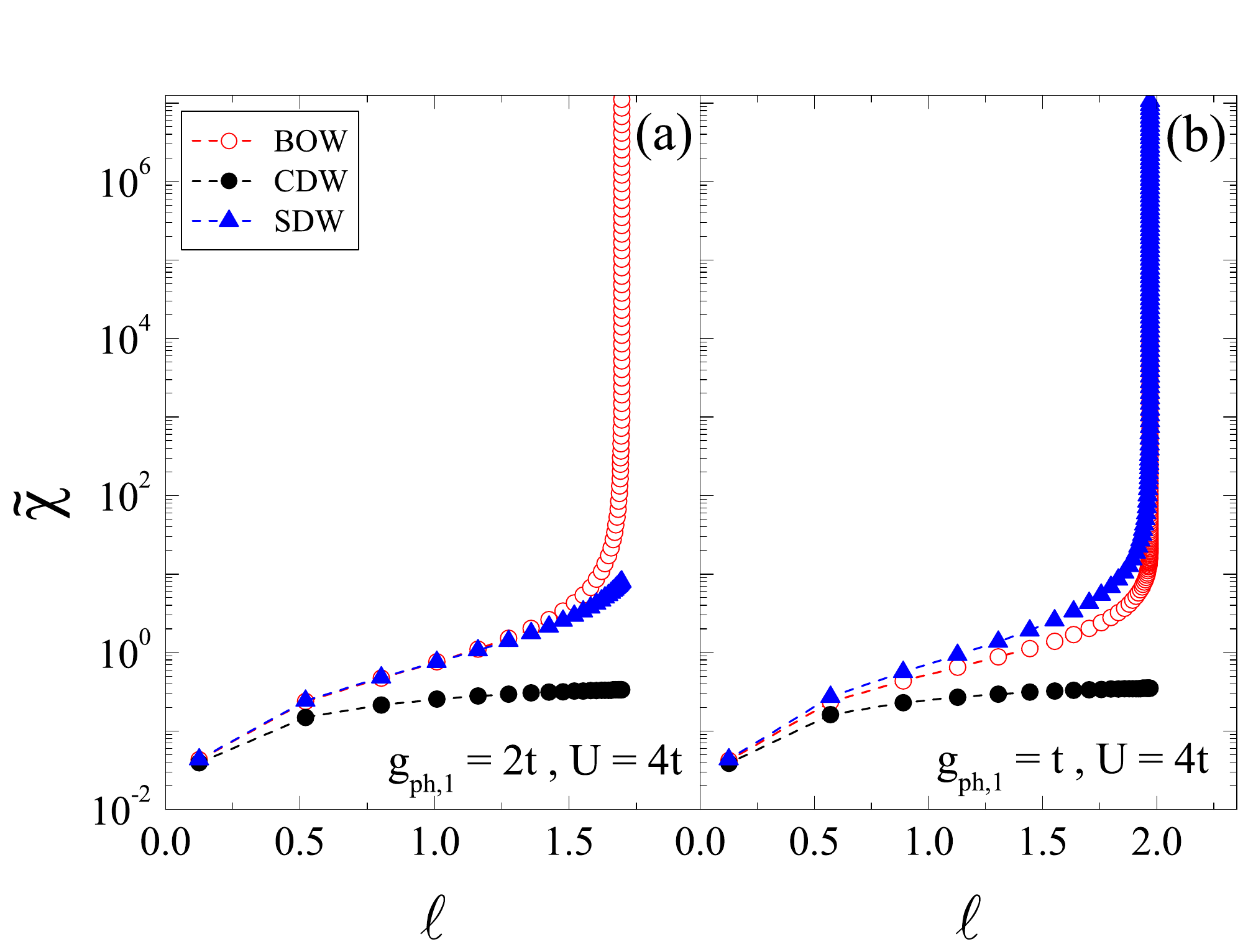}
\caption{\label{fig8} (Color online) Selected susceptibilities versus $\ell$ in   the  P-BOW (a) and M-SDW (b) parts of the phase diagram of the Peierls-Hubbard model of Fig.~\ref{fig9}~(a) at $\omega_D=t$. }
\end{figure}
\begin{figure}
 \includegraphics[height=11cm,width=8.0cm]{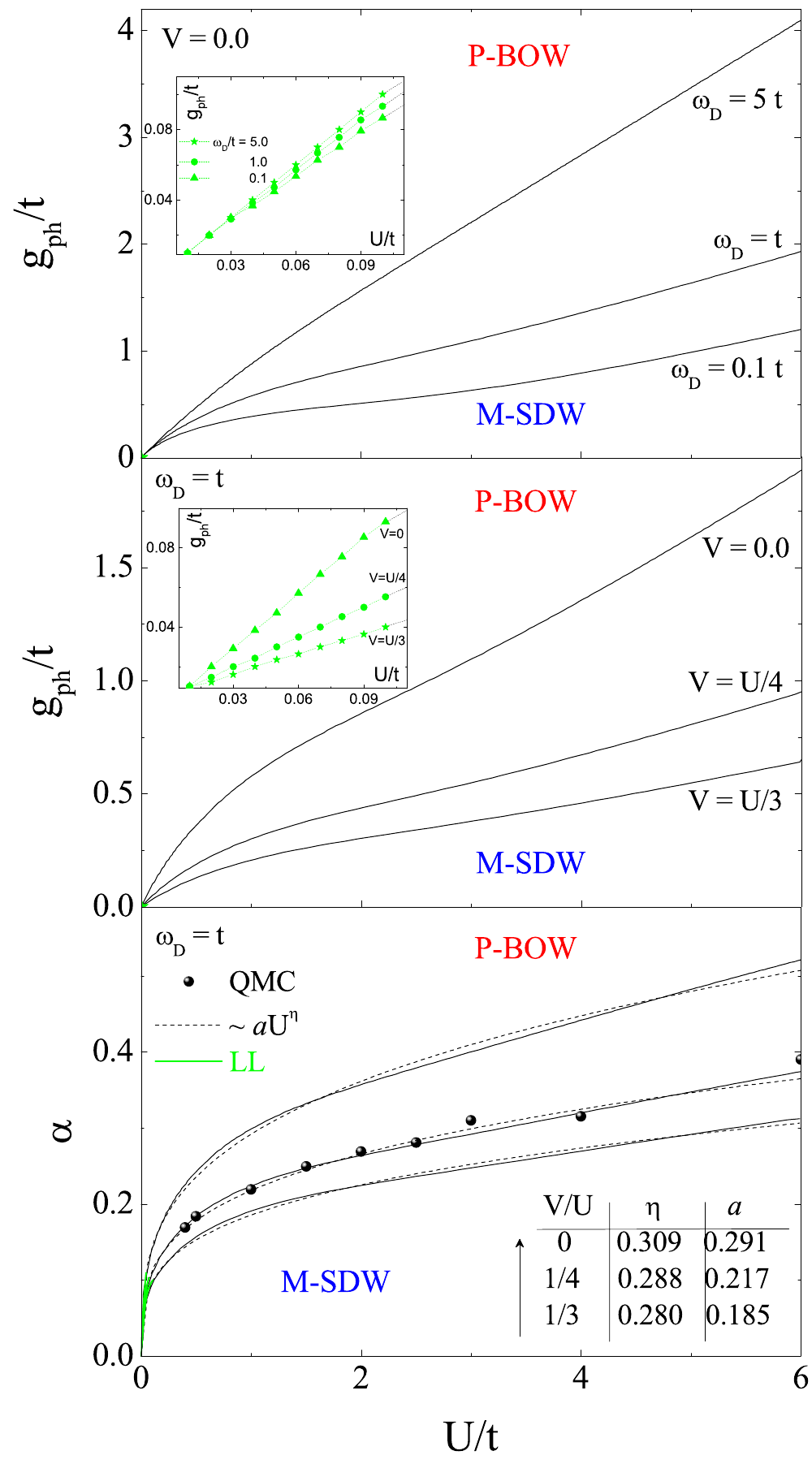}
\caption{\label{fig9} (Color online) Phase diagram   of   the PH model for  (top)  $\omega_{D}/t=0.1, 1,$ and $5$ at $V=0$; (middle) $\omega_D/t=1$ and $V\ge 0$  (inserts: zoom of the Luttinger liquid  line at very small couplings);   (bottom) comparison with a power law dependence of the boundary line between P-BOW and M-SDW (dashed lines) for different values of $V$ (bottom). The full circles are the QMC results of Ref.~\cite{Sengupta03}.}
\end{figure}

We now  examine the PH model at finite repulsive $U$. As   mentioned in Sec.~\ref{RG}, the coupling of electronic bond transfer to acoustic phonons    introduces particular initial conditions for the phonon-mediated  contribution to couplings of the PH model in (\ref{PH}). Only the phonon induced backscattering is attractive, whereas the umklapp term, like $U$, is repulsive.  According to (\ref{gBOW}), such a combination clearly favours the occurrence of the Peierls BOW (P-BOW) phase against CDW  whose  enhancement is totally absent across the $(U,g_{\rm ph})-$plane.    As shown in Fig.~\ref{fig8},    either a BOW or a SDW singularity is found at finite $g_{\rm ph}$, as a function of $U$; their tracking at $V=0$ yields the phase diagram   of Fig.~\ref{fig9} (top)  at different Debye frequencies. 

\begin{figure}
 \includegraphics[height=8cm,width=9.0cm]{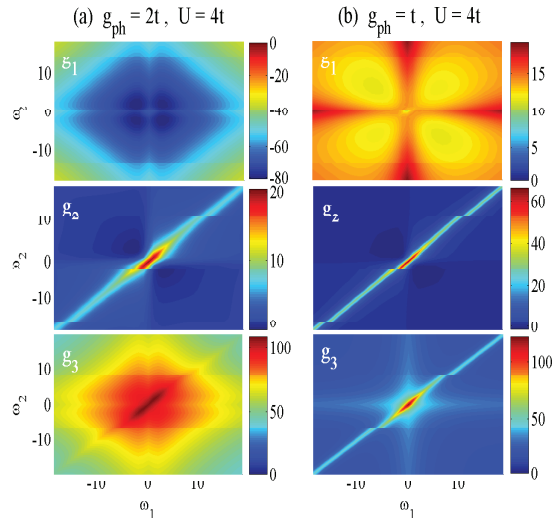}
\caption{\label{fig10} (Color online) Contour plot of the scattering amplitudes $g_i(\omega_1,\omega_2,\omega_1)$ in the $(\omega_1,\omega_2)$-plane for the Peierls Hubbard phase diagram of Fig.~\ref{fig9} (top) close to $\ell_c$  and at  $\omega_D=t$ : (a), the P-BOW phase   and (b), the M-SDW phase.}
\end{figure}

 On the P-BOW side, the  gap amplitude $\Delta = E_F e^{-\ell_c}$ extracted  at $\ell_c$ [Fig.~\ref{fig8} (a)], is always larger than the $U=0$ adiabatic mean-field value, $\Delta_0 =E_Fe^{-1/2\tilde{g}_{\rm ph}}$. Finite $U$ enhances  BOW correlations and then strengthens their coupling  to   2$k_F$ phonons in Eqs.~(\ref{g1}-\ref{g3})\cite{Caron84,Bindloss05}. The enhancement of the Peierls order parameter by repulsive $U$ is a well-known  result of early numerical simulations on that model~\cite{Hirsch83b,Mazumdar83}. Regarding the scattering amplitudes, singularity of  $g_1(0,0,0)$ is found  in the attractive sector,  while the strong-coupling peaks of both  $g_2(0,0,0)$ and umklapp $g_3(0,0,0)$ are for positive values (Fig.~\ref{fig10}). These   are indicative of a fully gapped phase, as expected for the commensurate Peierls order at half-filling.  The frequency dependence of the $g_i(\omega_1,\omega_2,\omega_1)$ displays similar features over the whole P-BOW region. However,  the influence of strong coupling  at finite frequency on  correlations remains weak.

 The  transition line between P-BOW and M-SDW states in the phase diagram is also of interest. Owing to the reinforcement of electronic BOW correlations by $U$, the transition to the M-SDW line occurs below the bisecting line $g_{\rm ph}=U$, in contrast to what has been obtained for the HH model (Fig.~\ref{fig2}).  As a result, moving toward   the adiabatic limit for small $\omega_D$ the P-BOW phase grows in importance to the detriment of M-SDW. In Fig.~\ref{fig9} (bottom),  we have plotted  the phonon coupling parameter $\alpha[\equiv (g_{\rm ph}\omega_D/8t^2)^{1/2}]$ versus $U$ for the transition line \cite{Sengupta03}. The trace   fits fairly  well the power-law  $\alpha_c \sim U^\eta$, with the exponent $\eta\simeq 0.31$ at $V=0$.  Similar algebraic variation has been reported by QMC simulations at finite $V$ \cite{Sengupta03} (see Sec.~\ref{V}). 
 For most $g_{\rm ph}$ values, the line $\alpha_c(U) $ shows a direct transition between the P-BOW and M-SDW  phases [Fig.~\ref{fig9} (top)],   except for very small $g_{\rm ph}$, where there is a finite but minute $U$ interval   in which LL  metallic conditions prevail (see insets of Fig.~\ref{fig9}).  As in the HH model,  a power-law dependences of the BOW susceptibility is found, as shown in Fig.~\ref{fig5}~(b). 
 Otherwise, due to the  fact that for the  PH model both the Coulomb and phonon-induced terms   contribute  positively to umklapp scattering [Eqs.~(\ref{PH}) and (\ref{UV})], the charge   sector is   gapped and the conditions for the emergence of  an intermediate metallic phase  are not  satisfied.

As to the M-SDW phase at sufficiently strong $U$, their characteristics are  similar to the one found for the HH model. This is confirmed by the frequency profiles of scattering amplitudes of Fig.~\ref{fig10}~(b) close to $\ell_c$, when compared to  those of Fig.~\ref{fig2}~(b).  The backscattering amplitude develop sizable repulsive values with a minimum  for $g_1(0,0,0)$, subordinated to the singular positive peaks  $g_2(0,0,0)$ and $g_3(0,0,0)$ at positive values. The Mott gap is associated with     the strongest singularity in $\chi_{\rm SDW}$ at $\ell_c$ closely followed by $\chi_{\rm BOW}$, as shown in Fig.~\ref{fig8}~(b).

\subsubsection{Finite V}
\label{V}
We close the section by examining the influence of small nearest-neighbor repulsion $V$ on the structure of the phase diagram of the PH model. In Fig.~\ref{fig9} (middle), the critical line $\alpha_c(U)$ is plotted for different $V$ ($<U/2)$. We first observe that  the  P-BOW  region increases in size  with $V$, as described   by the  fit  to the expression $\alpha_c = aU^\eta$ showing  a drop of the coefficient $a$ and a  power-law exponent $\eta$ which depends on the value of $V$. The reinforcement of BOW correlations and in turn of the Peierls state follows from the decrease (increase) of  $g_1$ ($g_2$) by $V$ in Eq.~(\ref{UV}), which compensates the drop in umklapp and then   according to (\ref{gBOW}) makes the BOW correlations and its coupling to phonons larger. The frequency dependence of the $g_i(\omega_1,\omega_2,\omega_1)$ projected in the $(\omega_1,\omega_2)$-plane present essentially the same characteristics as  those obtained in the  Hubbard case for $V=0$ (Fig.~\ref{fig10}).

The results obtained at  $V=U/4$ can be compared to QMC simulations performed in the same model conditions.\cite{Sengupta03} As shown in Fig.~\ref{fig9} (bottom), the power law  found by   QMC   for $\alpha_c$    is fairly  well reproduced  by the RG method over the whole   interval  of  acceptable couplings for a  weak coupling scheme.

\bigskip
\section{Summary and concluding remarks}

Summarizing,  we have studied the phase diagrams of  1D Holstein-Hubbard and Peierls-Hubbard models at half filling by a weak-coupling  RG method. The ($U$,$g_{\rm ph})$ phase diagrams that are mapped out from the susceptibilities and frequency-dependent scattering amplitudes at the one-loop level adhere for the most part  to the results of  numerical simulations,  in particular those of quantum Monte Carlo methods for which the most detailed studies have been carried out and the best comparison can be established in weak coupling. 

In the case of the Holstein-Hubbard model, the RG results  pinpoint a precise delimitation of an intermediate region separating the commensurate CDW and  a Mott-insulating SDW phases in the phase diagram.  The singular static superconducting correlations   found throughout this region, though always subordinate to CDW and tied to irrelevant  umklapp scattering at zero frequency, are congruent with the metallic character of this region. The present results   also revealed  an internal structure of  the phase  boundary with the M-SDW phase with the presence of distinct phases,  unanticipated and unexplored from the viewpoint of previous works on this model. A very weakly correlated Luttinger liquid  followed by a   BOW phase characterized by either a charge or spin gap has been found depending on the strength of the Coulomb term. The BOW phase replaces  the   Mott-SDW state as a result of  retardation, whose dynamics   generates  an attraction in the  backscattering amplitude alone, in analogy with momentum-dependent effects for the appearance of the BOW phase enfolding  the $U=2V$ line of the 1D extended  Hubbard  lattice model.

A similar analysis carried out on the Peierls-Hubbard model with its peculiar momentum- and frequency-dependent   scattering amplitudes   rather reveals  a Mott SDW state competing with the Peierls-BOW order  whose prominence   in the phase diagram grows as a function of retardation, and  intrasite and inter site Coulomb terms $U$ and $V$.  The nearly structureless boundary  between the  two ground states in the  $(U,g_{\rm ph})- $plane of the phase diagram  follows a power-law profile  compatible with numerical simulations.

\begin{acknowledgments}
C. B thanks  Martin Hohenadler for discussions on several aspects of this work. The financial support of the National Science and Engineering Research Council of Canada (NSERC) and the R\'eseau Qu\'eb\'ecois des Mat\'eriaux de Pointe (RQMP) is also acknowledged. Computational
resources were provided by RQCHP and Compute
Canada.
\end{acknowledgments}
 \bibliography{/Users/cbourbon/Dossiers/articles/Bibliographie/articlesII.bib}

\begin{thebibliography}{37}
\bibitem[{Not({\natexlab{b}})}]{NoteHH1}
\bibinfo{note}{In reference \cite{Bakrim07} where the same RG technique has
  been used for the purely Holstein limit at $U=0$, no singular superconducting
  correlations were found besides the dominant charge-density-wave ones. This
  was due to the range of couplings $g_{\rm ph}$ and frequency $\omega_0$ used
  in that work which lies above the intermediate region in the phase diagram.}
\end{thebibliography}
\end{document}